# Impact of material properties on the spectral broadening process in thin solid media

Viktor Pajer,[1,*] Levente Lehotai,[1] János Bohus,[1] Balázs Tari,[1] Mikhail Kalashnikov,[1] Ádám Börzsönyi,[1] and Roland S. Nagymihály[1]

[1]ELI ALPS, ELI-HU Non-Profit Ltd., Wolfgang Sandner utca 3., Szeged, H-6728, Hungary

**ABSTRACT**. The post-compression of high peak power laser pulses can efficiently increase the peak intensity on target without the need to build additional amplification stages and enlarge the compressor optics. However, to our best knowledge, the spectral broadening of such pulses with different nonlinear media has not yet been investigated systematically. Here we performed an experimental campaign with mJ-class 25 fs pulses, where the conditions of high energy post-compression were emulated. Fused silica, sapphire, and YAG plates were tested for spectral broadening and compressibility while measuring the spatio-spectral properties of the broadened pulses. We conducted a series of numerical simulations to have deeper insight on the broadening process, to better understand the experimental results and to find optimum parameters for future experiments.

## I. INTRODUCTION.

Thanks to the rapid evolution of high-energy laser systems in the past decades, the intensity in the focus reaches $10^{22}$–$10^{23}$ W/cm$^2$, due to which high-field research can be conducted with relatively compact laser-driven sources [1, 2]. To create more extreme conditions, the peak power of the driver pulses must be further increased [3, 4]. One possible solution to this end is boosting the energy in an additional amplifier, which is expensive, requires a lot of effort, raises many technical challenges and makes the operation and maintenance of the system more complex, especially at high repetition rates [5, 6]. An alternative, more compact approach is to reduce the pulse duration while preserving the energy content as much as possible. In high-energy Ti:Sa systems, the spectral bandwidth and consequently the pulse duration are still limited by the gain medium and the stretcher-compressor configuration, where the typical output is in the range of 25–30 fs [4, 6, 8]. Reaching sub-20 fs duration in a Ti:Sa based system is a challenging task [9]; a hybrid amplification scheme [10] has been recently proposed as a possibility, but modifying an existing and working system is not straightforward.

Pulse duration can be also reduced with the widely used post-compression method [11]. Here, the spectrum broadens after the compressor due to third-order nonlinear processes (e.g. self-phase modulation (SPM)) either in gas or in solid-state media. After the accumulated spectral phase is compensated by chirped mirrors, the pulse is recompressed to a duration below the initial value. Currently, post-compression is widely used in different laser systems with pulse energies not exceeding few tens of millijoules [11].

This technique has different schemes and implementations. In experiments using hollow core fibers or multipass cells, several studies have reported the optimal conditions for different input pulses through analytical and numerical calculations [12, 13]. However, the above-mentioned methods have severe limitations due to their geometry and mostly because of the damage threshold of the employed optical elements. Consequently, they are not suitable for high-energy laser systems.

On the other hand, the post-compression technique was proposed for high-energy pulses a few years ago, and significant spectral broadening was demonstrated by free propagation in solid thin plates [14-17]. One of the most important requirements of high-energy, and therefore big beam aperture post-compression is to achieve spatially homogeneous spectral broadening and, consequently, identical pulse duration across the entire beam cross-section. Obviously, the application of this method to high-energy laser systems has drawbacks, since during the nonlinear interaction process the laser intensity can be tuned in a limited range only, while modulations in the spatial intensity distribution are inherently present in the near field of high-energy laser systems.

The optimization of post-compression in high-energy laser systems requires a nonlinear medium with optimal properties that ensure significant broadening, while minimizing local self-focusing [18]. In experiments based on the thin plate approach, researchers mostly use fused silica or quartz as nonlinear material even at low energy levels, except for a few cases [19-22]. However, testing different materials or finding an optimal arrangement with a large, high-energy beam is a difficult and time-consuming task, as no detailed theoretical analysis has been carried out to determine the effect of material properties in case of the thin plate post-compression technique.

*Contact author: viktor.pajer@eli-alps.hu

We performed a series of experiments in a compact test environment where we mimicked the conditions of PW-class post-compression setups in terms of intensity and spatio-spectral distribution. We also present the results derived from numerical simulations about the expected spectral broadening, spatio-spectral behavior and optimal compression in a single thin plate arrangement.

## II. EXPERIMENTS
### A. Experimental setup

In our proof-of-principle experiment (Fig. 1 (a)) we used the HF-100 arm of the HFPW laser at ELI ALPS [6]. To match the conditions of a high-energy laser beam, we considered three main parameters which should be as close as possible to the real ones: pulse duration, flat-top beam profile and intensity on the thin plate. First, we shaped the output spectrum of HF-100 with an acousto-optic programmable gain filter (Mazzler, Fastlite) to reach a 25 fs compressed pulse duration. Then, we placed an apodizing reflective filter (NDY10A, Thorlabs) in front of the grating compressor. Here, with a 9 mm beam diameter, optimal for the apodizing filter, we could transform the Gaussian spatial intensity distribution into a sixth-order super-Gaussian one (Fig. 1 (b) and (c)). To avoid filter damage and to have the proper intensity on the nonlinear material, the energy was tuned with the help of an achromatic half-wave plate (AHWP10M-980, Thorlabs) and a polarizing beam splitter cube (CCM1-PBS252/M, Thorlabs) placed between the amplifiers and the filter. Finally, the output of the grating compressor was imaged with a demagnifying telescope (M = 0.13) onto the thin plate, resulting in a d = 1 mm beam size. As a starting point, we compared three different materials: fused silica (FS), single crystal YAG and polycrystalline sapphire with one, two and one millimeter thicknesses, respectively. To avoid any ionization or SPM in air, the second mirror of the telescope (M2) and the thin plate were placed in a small vacuum chamber with a footprint of 12×12 cm. The input and output windows were placed in Brewster angle. The dispersion of the input window was compensated with an acousto-optic programmable dispersive filter (Dazzler, Fastlite), through which an initial pulse duration of 25 fs was reached in front of the thin plate. As a collimated beam was propagating through the plate, broadening occurred in the near-field.

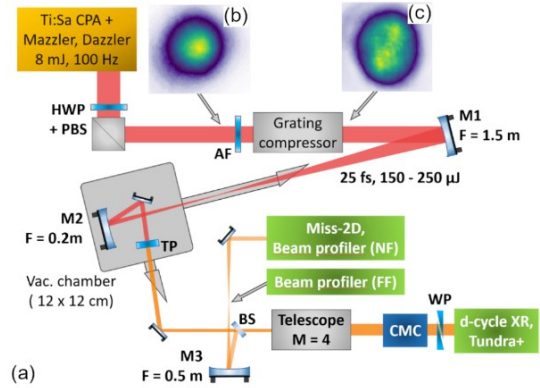

FIG. 1. (a) Schematic view of the experimental setup. HWP: half wave plate, PBS: polarizing beam splitter, AF: apodizing filter, TP: thin plate, BS: removable beam sampler, WP: wedge pair, CMC: chirped mirror compressor. M1 and M2 are curved mirrors used to image the output of the compressor on the thin plate. The M3 curved mirror imaged the plane of the plate onto the camera (NF) and onto the entrance slit of the 2D spectrometer. Insets: (b) beam profile before and (c) after the apodizing filter.

After coupling out the broadened pulse from the chamber, two arms could be used separately for diagnostics. In one arm, the beam was attenuated with a glass wedge, and the back surface of the nonlinear medium was imaged (M3 in Fig. 1 (a)) on a 2D spectrometer (MISS-L-B, Femto Easy) to measure the spatially resolved spectrum. At the same position, the near-field (NF) beam profile was recorded with a beam profiler camera (CinCam, CMOS-1.001-nano, Cinogy). In the focus of the M3 imaging mirror, we recorded the far-field (FF) beam profiles, too. After removing the wedge, in the second arm the beam was magnified fourfold by an afocal telescope and the pulses were compressed in a chirped mirror compressor (CMC in Fig. 1 (a)) consisting of seven pairs of chirped mirrors (PC1607, Ultrafast Innovations). An additional fused silica wedge pair was used for further optimization. The pulse duration was measured with a d-cycle XR (Sphere Ultrafast Photonics). To achieve a similar intensity distribution as in a high-energy setup, the input energy was set between 150 and 250 µJ. Consequently, the estimated intensity on the plates varied between 0.45 TW/cm$^2$ and 1.2 TW/cm$^2$. We note here that the spatial modulations drastically increased above 250 µJ and the signs of small-scale self-focusing and ionization appeared. The throughput of the setup was around 80%. The losses were mainly introduced by the uncoated surfaces of the thin plates as they were placed at 0° due to the limited geometry. The silver mirrors, which have ~97% reflection in this wavelength range, also reduced the efficiency. The energy stability after the setup was the same as the laser output with a typical value of 0.66% RMS and 4.22% PTP for one minute.

*Contact author: viktor.pajer@eli-alps.hu

## B. Compressibility

To achieve the broadest possible spectrum, the input energy was set to 250 µJ, except in the case of YAG, where it was reduced to 150 µJ. As the only available YAG plate was twice as thick as the other two and it has the highest nonlinear refractive index (Table I), we needed to lower the energy to achieve a similar bandwidth. Consequently, the estimated intensity on the plates would be 1.21 TW/cm$^2$ and 0.72 TW/cm$^2$, respectively. Fig. 2 (a) shows the spatially integrated input and broadened spectra for different materials. It is visible that in each case, we could reach a spectrum spanning from 700 to 940 nm at $10^{-2}$ intensity level despite the accumulated B integral was different. On the other hand, the bandwidth at $1/e^2$ level was 136 nm, 125 nm and 165 nm in case of FS, sapphire and YAG, respectively. In addition, the spectrum in the latter case was less modulated. This significant difference in the bandwidths appeared in the compressed pulse durations (Fig. 2 (b)) too, where the shortest pulse was 11.8 fs (YAG) while the narrowest spectrum supported 16 fs (sapphire). Altogether 560 fs$^2$ was compensated in the CMC, including approximately 268 fs$^2$ from the contribution of the 3 mm output window and the additional 5 m propagation in air (telescope and the CMC). The additional differences originating from material dispersion were fine-tuned by a wedge pair. We reached a maximum compression factor of two. It is very close to the theoretical value described in Section II.C, especially if we consider the complex structure of the spatio-spectral profile, including both the spatial and spectral modulations.

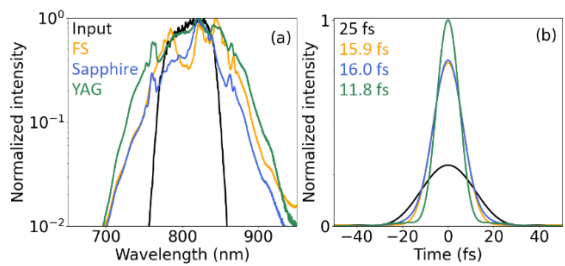

FIG. 2. (a) Spectra recorded before and after the vacuum chamber. (b) Measured temporal intensity profiles (FWHM pulse durations are indicated in the legend) after the CMC.

## C. Spatio-spectral homogeneity

As mentioned in the Introduction, the main requirement in high-energy post-compression processes is to achieve homogeneous spectral broadening. Thus, we recorded the spatially resolved spectrum at the image plane of the thin plates made of various materials. For the sake of simplicity, we present only the spatially resolved spectra measured in the vertical direction, noting that we observed very similar patterns on the perpendicular axis. In case of FS, the spectrum significantly narrows towards the beam edges and shows a more structured intensity distribution, especially around the central wavelength (Fig. 3 (a)). After the other two materials, the spatially resolved spectra showed a much more even intensity distribution.

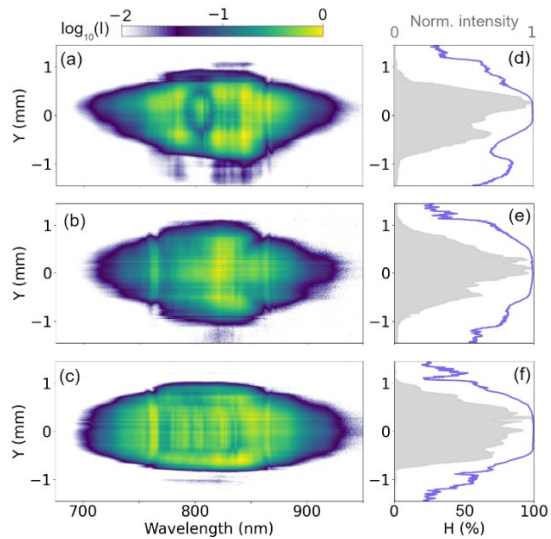

FIG. 3. Left: Spatially resolved spectra on a logarithmic scale, recorded in the vertical direction in case of (a) fused silica, (b) sapphire and (c) YAG. Right: The related spectral homogeneity (blue, [28]) and spectrally integrated beam profiles (gray filled). "I" indicates normalized intensity on the color bar.

To see whether the spatial intensity variations had any significant impact on the spatio-spectral distribution, we recorded the near- and far-field beam profiles of the broadened pulses. In addition, it was crucial to maintain good beam quality because any inhomogeneity in the spatial distribution can affect the interaction in the target chamber and can be detrimental to the beam transport optics. The input beam profile had a maximum intensity variation of 30%, which increased to the still tolerable level to 50%, 50% and 40% in case of FS, sapphire and YAG, respectively during nonlinear propagation (Fig. 4). Here we emphasize that a further energy increase resulted in a broader spectrum and shorter pulse duration. However, it caused even higher spatial intensity variations, due to small-scale self-focusing, hence we consider the above-mentioned input intensity values as a best effort. The near-field beam profiles show that we could still preserve the spatial quality, although differences in distribution are clearly visible after the various materials, which can be associated to the spatio-spectral homogeneity. The halo around the beam originates from the laser (see Fig. 1 (c)). The high-quality focal spots (Fig. 4 (d)-(f)) also suggest that no significant wavefront distortion occurred during nonlinear propagation.

*Contact author: viktor.pajer@eli-alps.hu

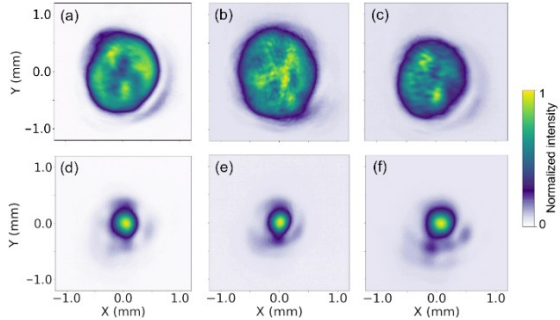

FIG. 4. Top: Near-field, normalized intensity beam profiles at the image plane of the thin plate in case of (a) fused silica, (b) sapphire and (c) YAG. Bottom: Related far-field beam profiles at the focus of the imaging mirror in case of (d) fused silica, (e) sapphire, (f) YAG.

## II. THEORETICAL MODEL

Both dispersion and Kerr type nonlinearity affect the nonlinear propagation of ultrashort pulses. Analogously to fiber optics, the dispersion length ($L_D$, Eq. (1)) and nonlinear length ($L_{nl}$, Eq. (2)) could be calculated as in equation 3.1.5 in [23]. As the effective mode area is not interpretable in case of a thin plate, we define $L_{nl}$ by using the intensity value. From a practical point of view, here we use the full width at half-maximum (FWHM) as the pulse width instead of the 1/e value.

$$L_D = \frac{\tau_0^2}{|GVD|} \quad (1)$$

$$L_{nl} = \frac{1}{\frac{n_2 \omega_c}{c} I_0} \quad (2)$$

Here, $\tau_0$ is the input pulse duration, $GVD$ is the group velocity dispersion at the central wavelength, $I_0$ is the peak intensity, $n_2$ is the nonlinear refractive index and $\omega_c$ is the central angular frequency. Based on these two characteristic lengths, one can predict which effect is dominant and what propagation behavior can be expected. In the normal dispersion regime, when the propagation length ($L$) is $L \ll L_D$ and $L \sim L_{nl}$ the pulse evolution is determined mainly by SPM, resulting in a broadened, modulated spectrum, while the temporal shape remains more or less unchanged. If $L \sim L_D$ but $L \ll L_{nl}$, the dominant effect is dispersion, which causes the temporal broadening of a transform-limited pulse. During nonlinear propagation, when both effects are present, initially it is mostly the spectrum that broadens. In parallel, dispersion continuously lengthens the pulse, the peak intensity decreases and, as a consequence, the spectral broadening gets saturated at a certain point.

Because of nonlinearity, the pulse duration increases even faster and the pulse shape starts to deviate from the original one, approaching a rectangular shape. At the same time, the modulation depth in the spectrum becomes less intense than expected by SPM only [23]. In addition, this happens faster in the central part of the beam where the intensity is the highest. After a certain length, the process slows down and the bandwidth and the accumulated B integral at the beam edges approach similar values as at the center.

### A. Spatio-spectral evolution

As the thickness of the available plates was limited, we ran a series of 2+1D simulations to better understand the evolution of the spatio-spectral behavior during nonlinear propagation. We applied the Sellmeier equations to take into account the material dispersion and we considered SPM, self-steepening and self-focusing when solving the Nonlinear Schrödinger equation (NLSE). As the seed, we used the measured spectrum and beam profile of the laser. Table I summarizes the main parameters of the materials [24-26] used for simulation and the related dispersion and nonlinear lengths.

TABLE I. Parameters of the nonlinear materials at 800 nm wavelength. n: refractive index, GVD: group velocity dispersion, $n_2$: nonlinear refractive index. $L_D$ and $L_{nl}$ are related to a 25 fs pulse with the indicated intensity.

| | Materials | Fused silica | Sapphire | YAG |
|---|---|---|---|---|
| | n | 1.45 | 1.76 | 1.82 |
| | GVD (fs$^2$/mm) | 36.16 | 58.04 | 98.45 |
| | $n_2$ ($10^{-16}$cm$^2$/W) | 2.0 | 3.1 | 6.3 |
| | $L_D$ | 20.16 | 12.57 | 7.44 |
| $L_{nl}$ | 1 TW/cm$^2$ | 0.55 | 0.41 | 0.21 |
| | 2 TW/cm$^2$ | 0.37 | 0.27 | 0.14 |
| | 3 TW/cm$^2$ | 0.28 | 0.21 | 0.11 |

Fig. 5 shows the simulated, spatially resolved spectra. The calculated, accumulated B integral is 2.4, 2.8 and 4.6 radians in fused silica, sapphire and YAG, respectively. The spatio-spectral modulations are more intense both in sapphire and in fused silica around the central wavelength. The differences are very similar but not as enhanced as in the experiments. The on-axis spectra clearly show, that after the YAG the spectrum is smoother than in the other two cases (Fig. 5 (f)) which is also in good agreement with the measurements. These results suggest that the material properties can indeed affect the spatio-spectral homogeneity, however the accumulated B integral can also play a role in the process.

*Contact author: viktor.pajer@eli-alps.hu

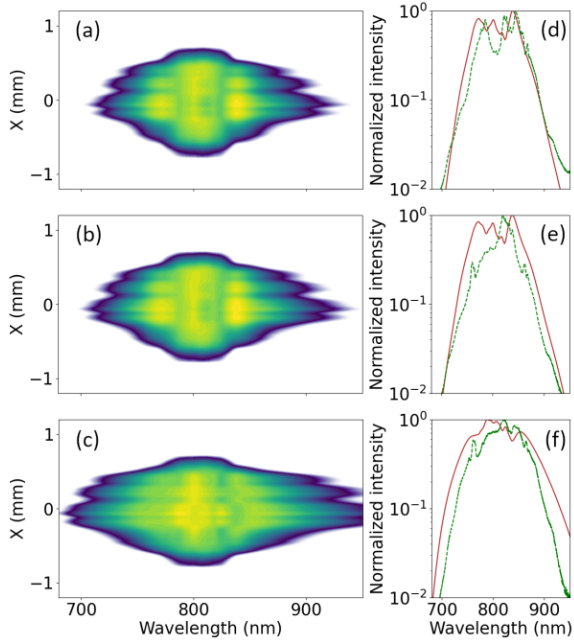

FIG. 5. Left: Simulated spatially resolved spectra on a logarithmic scale, using the measured spectrum and beam profile as input in case of (a) fused silica, (b) sapphire and (c) YAG. Right: The on-axis spectra (red, solid) with the corresponding measurement (green, dashed). The color map represents the same intensity levels as in Fig. 3.

To better understand the evolution of the spatio-spectral behavior during nonlinear propagation, we examined the spatially resolved spectrum after different propagation lengths. As the spatial and spectral modulations are unique for a real beam, we consider here a sixth-order super-Gaussian beam profile and 25 fs (full width half maximum) short Gaussian input pulse centered at 800 nm. Beam diameter is 1 mm to emulate the experimental conditions. To make the comparison easier, we introduce the normalized length, $z_0$ according to equation (3). $z_0$ is the propagation length, after which the pulse would be twice as long as the original when only GVD is considered.

$$z_0 = 0.5 L_D \qquad (3)$$

Fig. 6 shows the output after different nonlinear media and various propagation lengths. Here, we present the results at 1 TW/cm$^2$ input level which allowed us to obtain approximately the same B integral after each material. We emphasize here that the same propagation length, in terms of $z_0$, means different thickness for every nonlinear media. In case of fused silica, it is clearly visible that after $0.1z_0$ the intensity is already decreased at the central part (y=0 mm and λ=800 nm in Fig. 6 (a)) and this is more enhanced after $0.2z_0$ (Fig. 6 (b)). Considering the other two materials, broadening is very homogeneous in the first case and non-uniformity appears only after $0.2z_0$. On the other hand, further propagation results in a more even spatio-spectral distribution. This is evident if we consider that the pulse shape and consequently the spectrum starts to flatten due to the combination of dispersion and nonlinearity. In the beginning, this happens faster in the central part of the beam where the intensity is the highest. After a certain length, the process slows down and the bandwidth and the accumulated B integral at the beam edges approach similar values as at the center. The combination of these effects results in a relatively uniform spatio-spectral profile. As the $L_D$ and $L_{nl}$ of fused silica are the longest, SPM has a more dominant role at the beginning and longer propagation length is required to reach this regime. In case of the other materials, a shorter distance is sufficient to minimize the spatio-spectral modulations at the output. This can also explain why we obtained the most homogeneous broadening and a relatively smoother beam profile in YAG (Fig. 2). For the considered intensity level of ~1 TW/cm$^2$ the 2 mm thickness corresponds to around $0.5z_0$, while in case of the other two materials, 1 mm means less than $0.2z_0$.

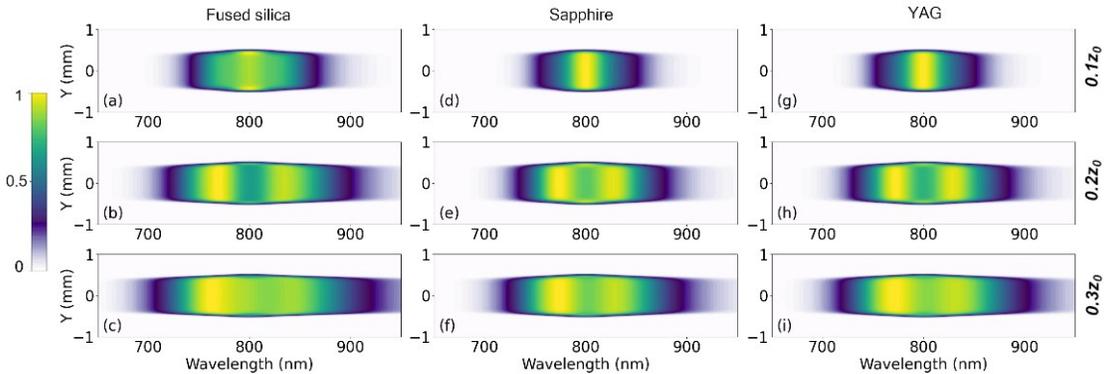

FIG. 6. Spatio-spectral intensity distribution after propagation in (a), (b), (c) fused silica; in (d), (e), (f) sapphire; and in (g), (h), (i) YAG. The corresponding normalized propagation lengths are indicated on the right side in each case.

*Contact author: viktor.pajer@eli-alps.hu

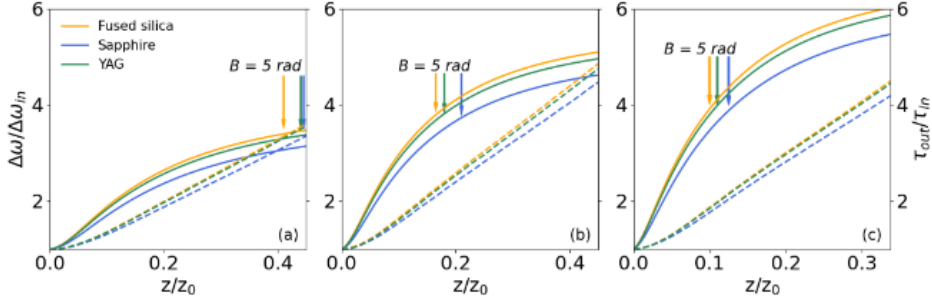

FIG. 7. Simulated spectral (solid lines, Δω/Δω$_{in}$) and temporal (dashed lines, τ$_{out}$/τ$_{in}$) broadening as a function of normalized propagation length (z/z$_0$) with different input intensities: (a) 1 TW/cm$^2$, (b) 2 TW/cm$^2$ and (c) 3 TW/cm$^2$. Arrows indicate where the accumulated B integral is 5 radians.

### B. Spectral broadening

To see how material properties affect the broadening process and compressibility in a thin plate post-compression setup, we simplified our numerical model and performed another series of one-dimensional simulations. We took the same parameters as before: 25 fs short Gaussian input pulse centered at 800 nm, propagating through a single thin plate. The nonlinear parameter used in the NLSE was set in a way to emulate 1, 2 and 3 TW/cm$^2$ input intensities. We applied the Sellmeier equations to take into account the material dispersion up to the fourth order and we considered SPM and self-steepening. In this section, we only discuss the spectral and temporal broadening, as well as the working conditions for optimal pulse compression based on the spectrum and temporal profile. We use RMS spectral bandwidth to describe the spectral broadening:

$$\Delta\omega = \sqrt{\frac{\int |E(\omega)|^2(\omega-\omega_c)^2 d\omega}{\int |E(\omega)|^2 d\omega}}, \quad (4)$$

where $\omega$ is the angular frequency, $\omega_c$ is the central angular frequency, and $|E(\omega)|^2$ is the spectral intensity. Fig. 7 shows the spectral and temporal broadening in different materials with various input intensities. In the last case (3 TW/cm$^2$), we limited the maximum propagation length at the point where the accumulated B integral exceeded 7 radians. As expected, higher intensity results in larger spectral bandwidths (solid curves) and there is no significant difference between the materials. On the other hand, 80% of the broadening occurs in the first quarter of propagation at each intensity level and it starts approaching saturation around 0.3z$_0$. In parallel, the pulse stretches continuously in time; at the end of the process, the pulse duration is 3 to 5 times longer than one would expect by GVD only.

### C. Optimal propagation length and compression factor

It is crucial to keep the temporal quality of the recompressed pulse. In other words, we need to find the propagation length after which the pulse quality is still acceptable, and a reasonable compression factor can be obtained. It is well known that SPM can generate or enhance spectral modulations, which lead to the appearance of temporal side-lobes and lower peak power than ideal after recompression. The highest peak power would be achieved as a compromise when the energy content of the side pulses is minimized, and the bandwidth is reasonably broadened. The best compression factor and the optimal propagation distance can be estimated by comparing the ratio of the peak power and the pulse duration of the compressed pulses to those of the input ones, which practically equals the energy content of the main peak. Hence, we define the pulse quality ($Q_p$) and compression factor ($K_c$) according to the following equations:

$$Q_p = \frac{P_C}{P_{in}}\frac{\tau_C}{\tau_{in}} \quad (5)$$

$$K_c = \frac{\tau_{in}}{\tau_c}, \quad (6)$$

where $P_c$, $P_{in}$, $\tau_c$, $\tau_{in}$ are the peak powers and the pulse durations of the compressed and input pulses, respectively. Post-compression setups include chirped mirrors, which mainly add group delay dispersion (GDD). To see how the compressed pulse duration evolves with the propagation length, we compensated the spectral phase by fitting and subtracting a second order polynomial, and calculated the compressed temporal profile at the output. Fig. 8 (a)-(c) shows that compressibility drops when the propagation length is short. Then it increases again with higher $z$ values until it reaches a maximum after a certain point. The corresponding propagation length ($z_{max}$), where we have the highest compression quality, becomes shorter with higher input intensity. In addition, the maximum achievable compression saturates shortly after $z_{max}$ (Fig. 8 (d)-(f)), contrary to what is expected from the spectral broadening (Fig. 7).

*Contact author: viktor.pajer@eli-alps.hu

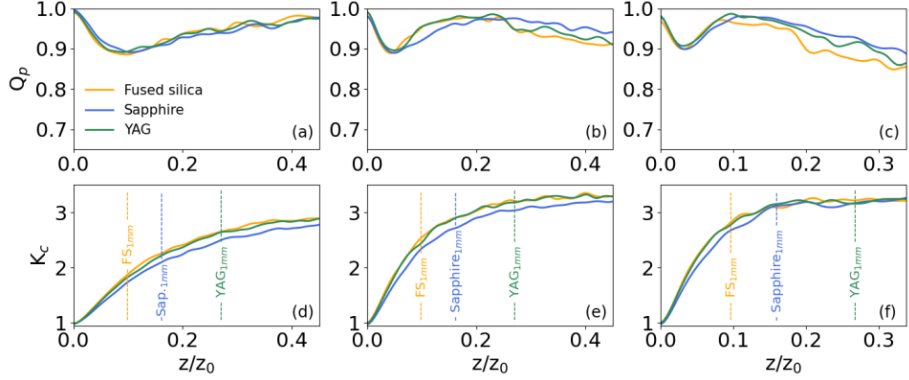

FIG. 8. Top row: Pulse quality as a function of normalized propagation length in case of (a) 1 TW/cm$^2$, (b) 2 TW/cm$^2$ and (c) 3 TW/cm$^2$ input intensity. Bottom row: The corresponding compression factor as a function of normalized propagation length in case of (d), 1 TW/cm$^2$, (e) 2 TW/cm$^2$ and (f) 3 TW/cm$^2$. The vertical dashed lines indicate the relevant values at the actual, 1 mm thickness.

These processes suggest that the shortest pulse duration would be around 8 fs independent of intensity or the choice of material in our case. We emphasize here that the transform-limited pulse duration gives a very similar upper limit regarding the compression factor. Besides the fact that the higher order phase components are not compensated in the simulation, the saturation of $K_c$ can be explained if we take into account that during propagation the pulse shape and consequently the spectrum gradually flattens as described previously. The higher the intensity, the faster it happens. The Fourier transform of such a super-Gaussian or rectangular signal will always be a Gaussian but with side-lobes. Thus, the compressed pulse quality inevitably degrades after a certain propagation length. We derived the optimal compression factor ($K_{c,opt}$) and propagation length from the pulse quality curves by simply determining $z_{max}$. According to reference [27], the compression factor for silica fiber can be approximated by $K_{c,opt} \approx f_c \sqrt{L_D/L_{nl}}$, which we found applicable to our case too. However, there are minor differences between the materials, the constant $f_c$ depends on the input intensity and it was found to be 0.48, 0.37 and 0.29 in case of 1, 2 and 3 TW/cm$^2$, respectively. Here, we note that except for the first case, these values are within the range where the maximum compression is already reached. For clarity, in Fig. 8 (d)-(f), the material thickness of 1 mm is also indicated for each material.

From a practical point of view, it is more beneficial to determine the optimal thickness of the plates ($L_{opt}$) instead of $z_{max}$. Thus, we extended the simulation series with lower intensity values and we interpolated $L_{opt}$. Fig. 9 shows the results as a function of input intensity. It is clearly visible that the favorable thickness decreases with higher $n_2$ (see also Table I) at lower intensities, but the difference becomes less significant above 1.5 TW/cm$^2$. We highlight that the highest quality at 3 TW/cm$^2$ would require a plate thinner than 0.5 mm. As it is unrealistic to produce one with a large aperture and high surface quality, pulse quality needs to be sacrificed under these conditions.

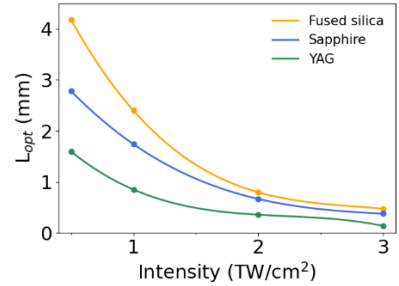

FIG. 9. Optimal plate thickness as a function of input intensity for various materials.

## CONCLUSION

In conclusion, we showed that the nonlinear propagation length is crucial to obtain spatially homogeneous broadening. Our results suggest that the achievable, compressed pulse duration is limited in a single thin plate post-compression setup. This minimum is independent of the material properties and of the input pulse intensity. We derived the optimal material thickness for 25 fs pulses at 800 nm wavelength with the highest quality which can help to optimize high-energy post-compression setups in advance.

## ACKNOWLEDGMENTS
The ELI ALPS project (GINOP-2.3.6-15-2015-00001); supported by the European Union and co-financed by the European Regional Development Fund.

## DATA AVAILABITY
Data underlying the results presented in this paper are not publicly available at this time but may be obtained from the authors upon reasonable request.

*Contact author: viktor.pajer@eli-alps.hu

# CONFLICT OF INTEREST
The authors declare no conflicts of interest.

*Contact author: viktor.pajer@eli-alps.hu